\documentclass[jgrga]{AGUTeX}

 \usepackage{lineno}

  \usepackage[]{graphicx}
  \setkeys{Gin}{draft=false}
\authorrunninghead{YEATES ET AL.}

\titlerunninghead{THE SUN'S OPEN MAGNETIC FLUX}

\authoraddr{J. A. Constable, School of Mathematics and Statistics, University of St. Andrews, St. Andrews, KY16 9SS, UK.}

\authoraddr{D. H. Mackay, School of Mathematics and Statistics, University of St. Andrews, St. Andrews, KY16 9SS, UK. (duncan@mcs.st-and.ac.uk)}

\authoraddr{A. A. van Ballegooijen, Harvard-Smithsonian Center for Astrophysics, 60 Garden Street, Cambridge, MA 02138, USA.}

\authoraddr{A. R. Yeates, Division of Mathematics, University of Dundee, Dundee, DD1 4HN, UK.  (anthony@maths.dundee.ac.uk)}

\begin{document}

%
%

\title{A Non-potential Model for the Sun's Open Magnetic Flux}
%

%
%

\authors{A. R. Yeates \altaffilmark{1,2}, D. H. Mackay \altaffilmark{3}, A. A. van Ballegooijen \altaffilmark{2}, and J. A. Constable \altaffilmark{3}}

\altaffiltext{1}{Division of Mathematics, University of Dundee, UK. }
\altaffiltext{2}{Previously at Harvard-Smithsonian Center for Astrophysics, USA.}
\altaffiltext{3}{School of Mathematics and Statistics, University of St. Andrews, UK. }

%
%


\begin{abstract}
Measurements of the interplanetary magnetic field (IMF) over several solar cycles do not agree with computed values of open magnetic flux from potential field extrapolations. The discrepancy becomes greater around solar maximum in each cycle, when the IMF can be twice as strong as predicted by the potential field model. Here we demonstrate that this discrepancy may be resolved by allowing for electric currents in the low corona (below $2.5 R_\odot$). We present a quasi-static numerical model of the large-scale coronal magnetic evolution, which systematically produces these currents through flux emergence and shearing by surface motions. The open flux is increased by $75\%$---$85\%$ at solar maximum, but only $25\%$ at solar minimum, bringing it in line with estimates from IMF measurements. The additional open flux in the non-potential model arises through inflation of the magnetic field by electric currents, with super-imposed fluctuations due to coronal mass ejections. The latter are modelled by the self-consistent ejection of twisted magnetic flux ropes.
\end{abstract}

%
%

%

\begin{article}

%
%

\newpage

\section{Introduction}

The ``open magnetic flux'' is the part of the Sun's magnetic field that is attached at one end to the solar photosphere but extends out to form the interplanetary magnetic field (IMF). It is of practical interest because it drives variations in the IMF near Earth. Short-term fluctuations are a source of geomagnetic activity \citep{pulkkinen2007}, while longer-term, secular variations---in particular, an apparent doubling over the last 100 years \citep{lockwood1999}---modulate the galactic cosmic ray flux \citep{beer2000} and are recorded in terrestrial records such as tree rings or ice cores. It has even been suggested that these variations may be linked to changes in global climate through variation of cloud cover \citep{svensmark1997}.

Empirical models are able to reproduce the long term variation of the open flux using either geomagnetic \citep{lockwood1999,lockwood2009e} or sunspot \citep{solanki2000,vieira2010} data as input. However, understanding the origin of the Sun's open magnetic flux requires knowledge of the structure and evolution of the magnetic field in the low solar corona. Unfortunately, direct magnetic measurements are available only at the photospheric level, so theoretical models are needed to extrapolate the magnetic field in the corona. The most popular approach has been to assume a potential field, {\it i.e.}, a vanishing electric current density. This gives a unique magnetic field solution in the region between an observed radial magnetic distribution on the photosphere $r=R_\odot$ and an outer ``source surface'' $r=R_{SS}$ where the field becomes purely radial. It is the magnetic flux through this source surface that determines the ``open flux'' in the model. Based on comparison of PFSS models with observed coronal structures, it is conventional to set $R_{SS}=2.5R_\odot$ \citep{schatten1969,altschuler1969}, although this leads to one criticism of the PFSS model: the real Alfv\'{e}n  radius, above which the solar wind kinetic energy density dominates the magnetic energy density, is of the order $10R_\odot$ \citep{marsch1984}. Nevertheless, PFSS extrapolations with $R_{SS}=2.5R_\odot$ are able to reproduce the locations of many He 10830\AA{} coronal hole boundaries \citep{wang1996,schrijver2003}, which map the distribution of at least some open field line footpoints on the solar surface. However, when the magnitude of open flux in such PFSS extrapolations is compared with {\it in situ} IMF measurements, it tends to be too low.

\begin{figure}
\noindent\includegraphics[width=20pc]{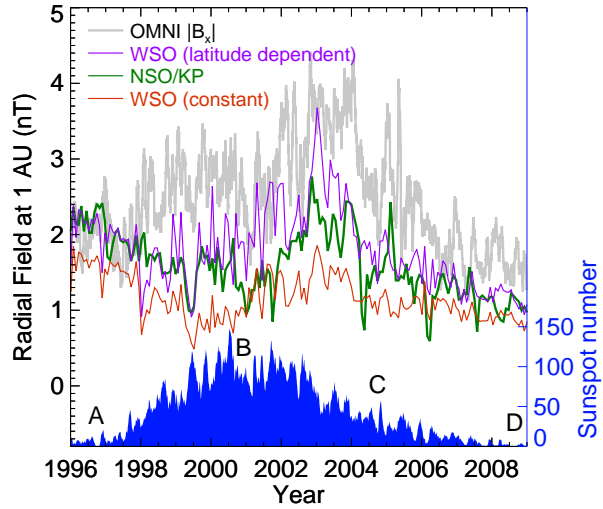}
\caption{Comparison of measured radial IMF at 1AU with various PFSS extrapolations over Solar Cycle 23. The thick grey line shows daily OMNI2 $|B_x|$ data with a two-stage averaging \citep[see][]{lockwood2009e}: an initial daily average of the signed data, to smooth out local small-scale fluctuations, followed by a 27-day running Carrington average of the unsigned data. Coloured lines show PFSS predictions taken from WSO with latitude-dependent correction (purple), WSO with constant correction (red), and NSO/Kitt Peak (green). The smoothed monthly sunspot number (from SIDC) is shown in blue, with letters A to D indicating times of our NP simulations.}
\label{fig:cycle}
\end{figure}

To compare the predicted open flux with {\it in situ} observations of the IMF strength, the modelled open flux is mapped out to $1\,\textrm{AU}$, and it is assumed that non-radial motions in the intervening space act to even out the distribution in latitude so that it becomes uniform \citep[e.g.,][]{wang2000}. This latter assumption is based on out-of-ecliptic magnetic field measurements by {\it Ulysses}, which show that the radial heliospheric flux is essentially independent of heliographic latitude \citep{balogh1995,smith2001}. The predicted radial field strength from the model at $1\,\textrm{AU}$, $B^E$, is then given simply by scaling the total unsigned flux through the outer model boundary, {\it i.e.}
\begin{linenomath*} 
\begin{equation}
B^E = \frac{R_{SS}^2}{4\pi R_E^2}\int\left|B_r(R_{SS},\theta,\phi)\right|d\Omega,
\label{eqn:be}
\end{equation}
\end{linenomath*} 
where $R_E$ the radius of Earth's orbit. Several authors have used observed photospheric magnetograms to produce time series of PFSS extrapolations, comparing $B^E$ with the observed  $B_x$ component of the IMF \citep{wang1995,wang2000,schrijver2003,schuessler2006}. In Figure \ref{fig:cycle}, the grey line shows the observed $B_x$ for Solar Cycle 23, while the three colored lines show PFSS predictions of $B^E$ using Equation \ref{eqn:be} with different photospheric magnetic field data (Section \ref{sec:poten}). Section \ref{sec:poten} discusses these PFSS models in more detail; the important point is that all three PFSS predictions are too low compared to the observations, particularly during more active phases of the Cycle. 

This shortfall has been noted over the past three solar cycles \citep[e.g.,][]{riley2007}, and several possible explanations have been proposed. Firstly, the open flux may be increased simply by lowering the source surface $R_{SS}$ in the PFSS model below $2.5R_\odot$, but this leads to poor agreement with coronal structures observed during eclipses \citep{altschuler1969}, and $R_{SS}$ should arguably be {\it larger} than $2.5R_\odot$, not smaller \citep{jiang2010}.  It is possible that the source surface is not spherical and its shape varies over the cycle \citep{schulz1978,riley2006}, but the effect of this on the open flux is presently unclear. Secondly, a modified version of the PFSS model, the so-called ``current-sheet source surface'' (CSSS) model, is found to give better agreement with the observed IMF \citep{zhao1995,schuessler2006}. The CSSS model adds an inner spherical ``cusp surface'' $r=R_{CS}$ beneath the outer source surface (which is moved out to about $10R_\odot$). Beneath $R_{CS}$ (typically about $1.8R_\odot$) the field remains potential, but currents are introduced between $R_{CS}$ and $R_{SS}$. This technique reproduces the latitude-independent field found by {\it Ulysses} as well as the heliospheric current sheet. However, the CSSS model has essentially three free parameters ($R_{CS}$, $R_{SS}$, and a parameter controlling the current distribution), which may be varied in order to match the open flux to the observed IMF \citep[{\it e.g.,}][]{jiang2010}. A third possible explanation for the PFSS shortfall in open flux, proposed by \citet{fisk2006}, is that reconnection of open field lines with closed coronal loops could lead to a ``diffusion'' of open flux into otherwise closed regions, producing an additional component of open flux not included in PFSS models. However, this has not been demonstrated by observations \citep{rust2008}. A fourth possibility, recently suggested by \citet{lockwood2009d,lockwood2009e}, is that flow speed variations in the solar wind above $2.5R_\odot$ could lead to an increase in the unsigned radial flux at increasing distance from the Sun. Such a ``flux excess'' effect appears to be supported by observations from various spacecraft out to $20\,\textrm{AU}$, and could potentially also explain the difference between PFSS flux at $2.5R_\odot$ and IMF measurements.

In this paper, we suggest another possible explanation for the PFSS shortfall: the influence of electric currents in the low corona below $2.5R_\odot$. Such currents are expressly prohibited in the PFSS model, yet we may infer their presence throughout the solar cycle, both in newly-emerged active regions, {\it e.g.}, in X-ray sigmoids \citep{canfield1999,canfield2007}, and outside active latitudes, including long-lived H$\alpha$ filament channels \citep{martin1994,mackay2008} and coronal magnetic flux ropes \citep{gibson2006}. There have already been indications from global full-MHD models that allowing such currents will increase the open flux \citep{riley2006}. However, these models tend to assume an initial potential field which is then relaxed to equilibrium with an imposed solar wind structure, so do not take into account the time-dependent development of currents due to the emergence and interaction of active regions in the lower corona. In Section \ref{sec:model}, we describe an alternative, simplified global magnetic field model that is less computationally demanding yet takes these time dependent non-potential effects into account. The remainder of the paper applies this model to study the magnitude and variation of the open flux. In Section \ref{sec:outflow} we consider the effect of the outer boundary condition on the open flux, before showing in Section \ref{sec:np} how the NP model predicts enhancement of the open flux due to inflation of the magnetic field by coronal currents. Conclusions are given in Section \ref{sec:conclusion}.

\section{Open Flux in the PFSS Model} \label{sec:poten}

With a fixed source surface radius $R_{SS}$, the open flux in a PFSS extrapolation depends purely on the photospheric radial magnetic field at a single instant. This is the cumulative product of many  bipolar active regions which have emerged and been dispersed by surface flux transport. In fact, the lowest order multipole components of the photospheric field determine the open flux, since higher-order multipoles decay more rapidly with height \citep{wang2002,mackay2002}. For a few Carrington rotations, newly-emerged bipoles lead to fluctuations of the non-axisymmetric, equatorial dipole component. Later, however, this contribution decays, or cancels between neighbouring bipoles, and the net asymptotic contribution is to the axisymmetric dipole component. During a single Solar Cycle, there is a factor 2 modulation of the observed IMF ({\it e.g.}, Figure \ref{fig:cycle}), though the variation is typically only $10\%$--$20\%$ that of the closed flux \citep{wang2000}. The observed open flux is not quite in phase with sunspot activity (Figure \ref{fig:cycle}), peaking about 1--2 years after Maximum. Earlier PFSS extrapolations from surface flux transport models had trouble reproducing this observed phase relation \citep{mackay2002b}, although this might be explained either by the properties of active regions used \citep{schuessler2006}), or by including the contribution of CMEs, discussed in Section \ref{sec:ejections}.

Because it is determined by the photospheric magnetogram input, the open flux predicted by PFSS models depends on the source of these observational data, leading to differing predictions using magnetograms from different observatories. For observations made in the Fe 5250\AA{} line at either Mount Wilson Observatory (MWO) or Wilcox Solar Observatory (WSO), a major factor affecting the level of open flux predicted is the saturation correction applied to the magnetograph signal. To illustrate this point, we have computed PFSS extrapolations from WSO data with two alternative correction factors that have been used in the literature. The red line in Figure \ref{fig:cycle} uses a constant factor $1.85$, which is appropriate for the WSO instrument \citep[see][]{riley2007}, whereas the purple line uses a latitude-dependent factor $4.5 - 2.5\sin^2\lambda$ that was derived from MWO measurements of the same line \citep{ulrich1992,wang1995}. Clearly the latitude-dependent correction factor gives closer agreement to the observed $B^E$, even though it is arguably not applicable to the WSO instrument \citep{riley2007}. For comparison, the green line shows the PFSS prediction using alternative magnetogram data from NSO/Kitt Peak derived from the Fe 8688\AA{} line, where a similar saturation factor is not required \citep{arge2002}. This higher resolution data is used to derive the emerging bipole data input to our non-potential simulations.

It is clear from Figure \ref{fig:cycle}, however, that no matter which data are used, PFSS extrapolations predict too low an open flux over much of the cycle, particularly in more active epochs (such as the years 2000 or 2004). Agreement is better in Minimum periods, although more so in 1996 than in 2008. The non-potential enhancement described in this paper is one possible explanation for this solar activity-dependent discrepancy.

\section{Non-potential Evolution Model} \label{sec:model}

With the aim of moving beyond the PFSS model and understanding the evolution of currents in the coronal magnetic field, we have developed global, non-potential simulations based on the coupled flux transport and magneto-frictional model of \citet{vanballegooijen2000}. This non-potential (hereafter NP) model follows the time evolution of the large-scale magnetic field through a quasi-static sequence of near force-free equilibria, in response to flux emergence and shearing by surface motions. Currents are generated in the corona, and subsequent flux cancellation above polarity inversion lines tends to concentrate the associated magnetic helicity into either sheared arcades or twisted magnetic flux ropes \citep[through the mechanism described by][]{vanballegooijen1989}. The model in its present form has been successfully applied to study the formation of filament channels \citep{mackay2000,yeates2008a} and the initiation of coronal mass ejections \citep{yeates2009b}. The latter occur in the model when flux ropes grow too strong to remain in equilibrium \citep{forbes2000} and are ejected through the outer boundary of the numerical domain.

The NP model differs from PFSS extrapolations in a number of important respects. Firstly, the latter assume a current-free corona, whereas in the NP model currents are generated both by the emergence of active regions and by subsequent shearing by photospheric motions. The effect of these currents on the open flux is to increase it relative to a potential field. Secondly, 
the PFSS model uses a sequence of independent extrapolations, usually only once per solar rotation ($27.27$ days), with no information on how the coronal or photospheric fields have evolved. In contrast, the NP model follows the continuous evolution so that the magnetic field can retain a ``memory'' of previous interactions and topology. Because an observed magnetogram of the whole solar surface can be taken only once per rotation, we do not use synoptic magnetograms directly as input into the NP model, but rather use them to determine where on the solar surface new active regions have recently emerged. Corresponding magnetic bipoles are then inserted individually into the simulation to produce a continuously evolving boundary condition. A further difference is at the outer boundary.  To facilitate direct comparison with PFSS extrapolations, this boundary is kept at $r=2.5R_\odot$ in this paper. However, the ``source surface'' condition used in PFSS extrapolations, namely that the transverse magnetic field components vanish, can not be applied in the NP model since it would prevent the ejection of magnetic flux ropes through this boundary following their loss of equilibrium. Instead, we impose a radial outflow to simulate the effect of the solar wind in radially opening field lines that reach near to the boundary \citep{mackay2006}. This is discussed in Section \ref{sec:outflow}.

\begin{figure*}
\begin{center}
\noindent\includegraphics[width=34pc]{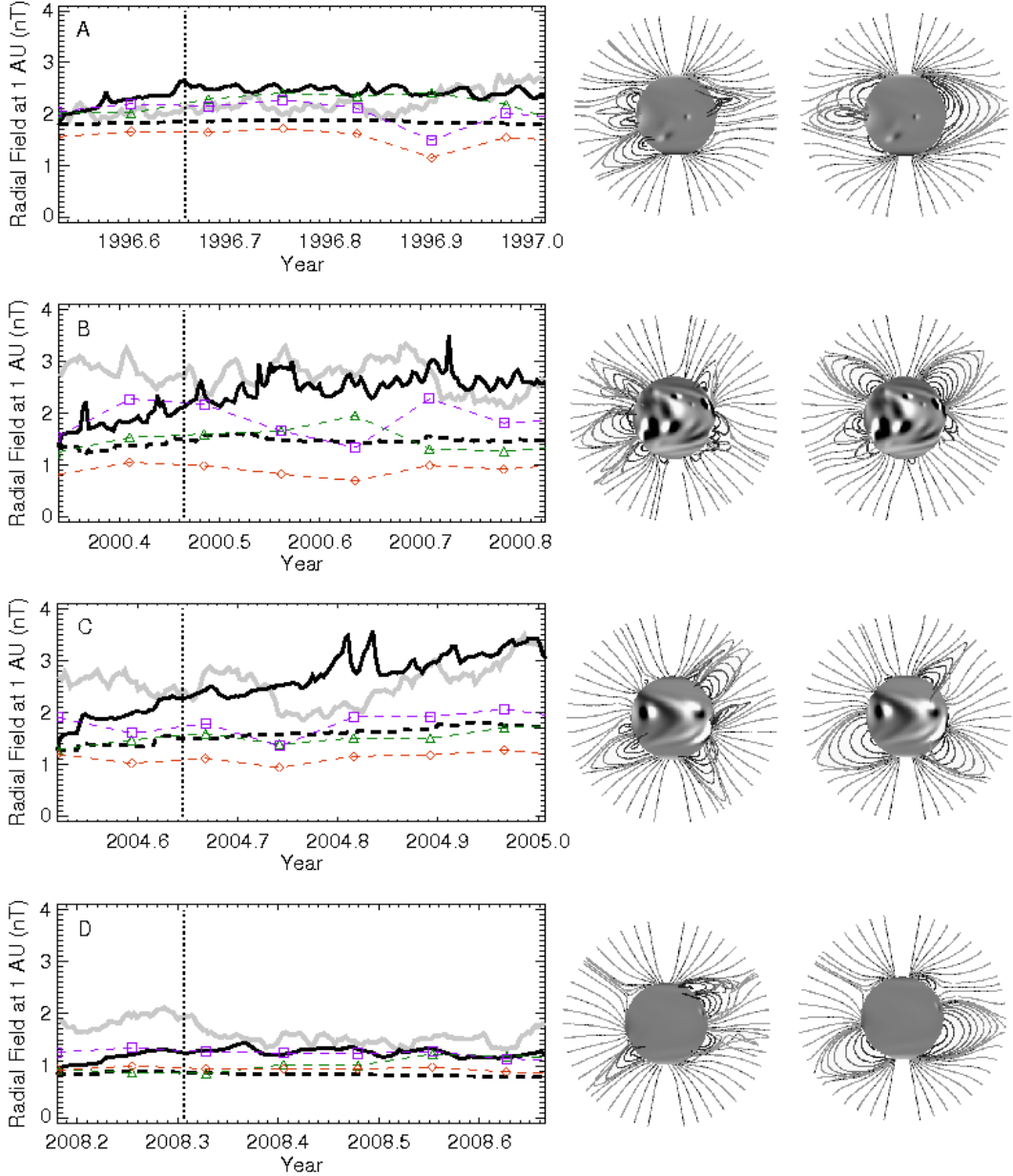}
\caption{Comparison of the NP model and PFSS extrapolations, for periods A to D. The plots on the left show the predicted $B^E$ and observed $|B_x|$, while the snapshots show the magnetic field in the NP model (middle) alongside a PFSS extrapolation (right) from the same radial photospheric field. In the left-hand plots, as in Figure \ref{fig:cycle}, grey curves show the observed $|B_x|$, thick black curves show predictions from the NP simulations, the black dashed line shows the PFSS prediction from the simulated photospheric field, and coloured symbols joined with dashed lines show PFSS predictions from observed magnetograms. The vertical dotted lines indicate the approximate end of the initial ``ramp-up'' period. In the snapshots, grey-shading shows radial magnetic field on the photosphere (white positive, black negative, with a saturation level of   $20\,\textrm{G}$ or $2\,\textrm{mT}$), while selected coronal magnetic field lines are traced from the plane of the sky.}
\end{center}
\label{fig:np}
\end{figure*}

\subsection{Model Equations}

The simulations use a domain extending from $0^\circ$ to $360^\circ$ in longitude, $-80^\circ$ to $80^\circ$ in latitude, and $R_\odot$ to $2.5R_\odot$ in radius \citep{yeates2008a}. Note that, to avoid the computational difficulties of grid convergence and coordinate singularity, our domain omits the regions within $10^\circ$ of the poles. These omitted regions amount to only $1.5\%$ of the solar surface. Although they may carry a larger fraction of the open flux at solar minimum, this is partially offset by displacement of flux to lower latitudes, and may well be within the observational uncertainty of the magnetograms at these high latitudes.

The large-scale {\it mean} magnetic field is evolved by the induction equation
\begin{linenomath*} 
\begin{equation}
\frac{\partial\mathbf{A_0}}{\partial t} = \mathbf{v}_0\times\mathbf{B}_0 + \mathcal{E}_0,
\label{eqn:induction}
\end{equation}
\end{linenomath*} 
where $\mathbf{A}_0(\mathbf{r},t)$ is the vector potential for the mean magnetic field $\mathbf{B}_0 = \nabla\times\mathbf{A}_0$, with gauge chosen to cancel the additional gradient term on the right-hand side of (\ref{eqn:induction}), and $\mathbf{v}_0(\mathbf{r},t)$ is the mean plasma velocity. The mean electromotive force $\mathcal{E}_0(\mathbf{r},t)$ describes the effect of small-scale fluctuations which are not resolved in our mean field model, such as braiding and current sheets produced by interaction with convective flows in the photosphere. We assume the form
\begin{linenomath*} 
\begin{equation}
\mathcal{E}_0 = -\eta\mathbf{j}_0,
\end{equation}
\end{linenomath*} 
where $\mathbf{j}_0=\nabla\times\mathbf{B}_0$ is the current density, and
\begin{linenomath*} 
\begin{equation}
\eta = \eta_0\left( 1 + 0.2\frac{|\mathbf{j}_0|}{|\mathbf{B}_0|}\right)
\end{equation} 
\end{linenomath*} 
is a turbulent diffusivity. The first term is a uniform background value $\eta_0=45\,\textrm{km}^2\,\textrm{s}^{-1}$ and the second term is an enhancement in regions of strong current density \citep{mackay2006}, introduced to limit the twist in helical flux ropes to about one turn, as observed in filaments \citep{su2009}.

Rather than solving the full MHD equations for the velocity $\mathbf{v}_0$, we approximate the momentum equation in the coronal volume by the magneto-frictional method \citep{yang1986}, setting
\begin{linenomath*} 
\begin{equation}
\mathbf{v}_0 = \frac{1}{\nu}\frac{\mathbf{j}_0\times\mathbf{B}_0}{B^2}.
\label{eqn:v}
\end{equation}
\end{linenomath*} 
This artificial velocity models the coronal field evolution through a sequence of near (nonlinear) force-free equilibria. In addition, a term is added to the radial component of $\mathbf{v}_0$ near the outer boundary; this will be described in Section \ref{sec:outflow}.

On the lower, photospheric boundary, the magneto-frictional velocity is not applied, but rather the radial magnetic field, $B_{0r}$, is evolved using the standard surface flux transport model \citep[][and references therein]{sheeley2005}. This surface component of our simulation was described in detail in \citet{yeates2007}. In terms of the vector potential $\mathbf{A}_0$ and standard spherical polar coordinates $(r,\theta,\phi)$, the evolution equations on the lower boundary $r=R_\odot$ are
\begin{linenomath*} 
\begin{eqnarray}
\frac{\partial A_{0\theta}}{\partial t} &=& u_\phi B_{0r} - \frac{D}{R_\odot\sin\theta}\frac{\partial B_{0r}}{\partial \phi},\label{eqn:surface1}\\
\frac{\partial A_{0\phi}}{\partial t} &=& -u_\theta B_{0r} + \frac{D}{R_\odot}\frac{\partial B_{0r}}{\partial \theta}.\label{eqn:surface2}
\end{eqnarray}
\end{linenomath*} 
Here $D=450\,\textrm{km}^2\,\textrm{s}^{-1}$ is a diffusivity modelling the random walk of magnetic flux owing to the changing supergranular convection pattern. The differential rotation velocity $u_\phi = \Omega(\theta)R_\odot\sin\theta$ uses the observationally-determined \citet{snodgrass1983} profile
\begin{linenomath*} 
\begin{equation}
\Omega(\theta) = 0.18 - 2.3\cos^2\theta - 1.62\cos^4\theta\,\textrm{deg}\,\textrm{day}^{-1},
\end{equation}
written in the Carrington frame. We include a poleward bulk flow, or meridional circulation
\end{linenomath*} 
\begin{linenomath*} 
\begin{equation}
u_\theta(\theta) = C\cos\left[\frac{\pi\left(\theta_\textrm{\small max} + \theta_\textrm{\small min} - 2\theta\right)}{2\left(\theta_\textrm{\small max} - \theta_\textrm{\small min}\right)}\right]\cos\theta,
\end{equation}
\end{linenomath*} 
as in \citet{mackay2006}. The constant $C$ is chosen to give a peak flow at mid-latitudes of $16\,\textrm{m}\,\textrm{s}^{-1}$, and the flow vanishes at the boundaries $\theta_\textrm{\small min}=10^\circ$ and $\theta_\textrm{\small max}=170^\circ$ of the computational domain.

\subsection{Observational Input}

Newly emerging magnetic bipoles are inserted based on active regions observed in synoptic normal-component magnetograms from the US National Solar Observatory (NSO) at Kitt Peak. The locations of these regions on the solar surface are determined from the magnetograms, but the exact date of emergence cannot be determined from the available observations, since at any time these show only the earthward face of the Sun. We simply insert all new regions 7 days before their date of first observation at central meridian, with their properties appropriately scaled \citep[as in][]{yeates2007}. Varying this time period does not change the accuracy of the surface magnetic field, and does not affect the level of the open flux on timescales of a solar rotation or longer \citep{schrijver2003}. The insertion is instantaneous, but is preceded by an artificial ``sweeping'' of strong magnetic field out of the insertion region, as described in \citet{yeates2008a}. This models the expected distortion of preexisting coronal field by a newly-emerging flux tube, and prevents the formation of disconnected flux in the corona. We find that this technique of inserting individual magnetic bipoles to model emerging active regions is able to retain high accuracy between the simulated and observed surface fields over many months \citep{yeates2007}.

The emerging bipoles take the mathematical form given in \citet[][equations 6 to 9]{yeates2008a}, with properties chosen to match the size, tilt, and magnetic flux of the observed regions. They are inserted in 3D, and are given a non-zero twist (magnetic helicity) through the parameter $\beta$, as described in \citet{yeates2008a}. The magnitude and, especially, sign of this emerging bipole helicity have been shown to influence both the chirality of filament channels \citep{mackay2005,yeates2008a} and the ejection rate of magnetic flux ropes \citep{yeates2009b}. Unfortunately, the optimum value of the helicity parameter for each bipole is poorly constrained by present-day observations, and must be arbitrarily chosen. The observations that do exist indicate a variation in both magnitude and sign of helicity both within individual active regions and between different regions, as well as a negative gradient of helicity with latitude \citep{pevtsov1995}. To approximate these features, we select a random $\beta$ value for each inserted bipole (up to a maximum $|\beta|=1$) from a normal distribution with mean $-0.4\lambda_0^\circ/25^\circ$ and standard deviation $0.4$, where $\lambda_0$ is the central latitude of the bipole.

\subsection{Four Simulations over Cycle 23}
 
In this paper we focus on four main simulation runs A to D, each covering $5.5$ months at different epochs of Cycle 23. The dates are shown in Table \ref{tab:periods}, and the periods are indicated in Figure \ref{fig:cycle}. The fourth column of Table \ref{tab:periods} shows the total number of bipoles emerged in each simulation run, as determined from the observed magnetograms. The right-most column indicates the instrument used; the older vacuum telescope (KPVT) was replaced in 2003 by SOLIS (Synoptic Optical Long-term Investigations of the Sun). The four runs A to D are illustrated in Figure \ref{fig:np}, where plots in the left column compare the modelled $B^E$ from both the PFSS and NP models with the observed IMF $|B_x|$. The snapshots in the middle and right columns show the NP simulation (left) and a PFSS extrapolation from the same, simulated, surface radial field (right). The black dashed lines in the left column of Figure \ref{fig:np} show $B^E$ for daily sequences of such PFSS extrapolations. Their agreement with the PFSS extrapolations from observed magnetograms (once per solar rotation; colored dashed lines) indicates that our technique for incorporating observed bipoles and simulating the surface field is sufficiently accurate to study the open flux.

\begin{table*}
\caption{Simulation periods.}
\label{tab:periods}
\begin{tabular}{lllll}
\hline
Period & Carrington Rotations & Dates & Bipoles & Instrument\\
\hline
A (Minimum) & 1911.5 to 1917 & 12 Jul 1996 to 04 Jan 1997 & 16 & KPVT\\
B (Maximum) & 1962.5 to 1968 & 03 May 2000 to 27 Oct 2000 & 122 & KPVT\\
C (Declining phase) & 2018.5 to 2024 & 08 Jul 2004 to 02 Jan 2005 & 52 & SOLIS\\
D (Minimum) & 2067.5 to 2073 & 06 Mar 2008 to 30 Aug 2008 & 14 & SOLIS\\
\hline
\end{tabular}
\end{table*}

It is evident from Figure \ref{fig:np} that the NP model for each period, after an initial ramp-up over about 1.5 months (indicated by the vertical dotted lines), settles on a level of $B^E$ which more closely matches IMF observations than that obtained from PFSS extrapolations. The initial ramp-up (to the left of the vertical dashed lines) demonstrates the importance of time evolution. This is the time taken for newly emerging bipoles and surface motions to evolve the coronal field away from the initial condition (a PFSS extrapolation) to some sort of equilibrium level of non-potentiality. To understand how the NP model achieves the observed level of open flux we must consider several contributory effects.

\section{Effect of the Outer Boundary Condition} \label{sec:outflow}

In our NP model, a radial outflow is imposed at the outer boundary $r=2.5R_\odot$ to model the effect of the solar wind in opening up coronal field lines above this height. A term of the form
\begin{linenomath*} 
\begin{equation}
\mathbf{v}_\textrm{out} = v_0\textrm{e}^{-(2.5R_\odot - r)/r_w}\hat{\mathbf{r}}
\end{equation}
\end{linenomath*} 
is added to Equation \ref{eqn:v}, where the exponential fall off (width $r_w=0.2R_\odot$) ensures that only structures near to the outer boundary are affected. Typically we set $v_0=100\,\textrm{km}\,\textrm{s}^{-1}$ as a compromise between a realistic solar wind speed and a reasonable time step in the integration (at least $30\,\textrm{s}$).

\begin{figure}
\noindent\includegraphics[width=20pc]{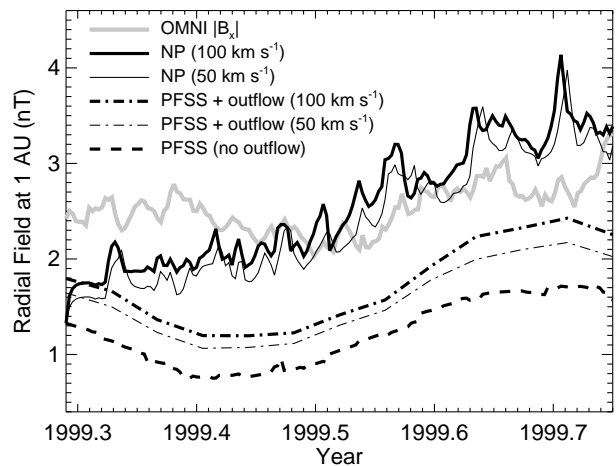}
\caption{Effect of a radial outflow at the outer model boundary, illustrated for a period in 1999. Thick lines show the predicted $B^E$ in simulations with $v_0=100\,\textrm{km}\,\textrm{s}^{-1}$, while thin lines show that for $v_0=50\,\textrm{km}\,\textrm{s}^{-1}$. The two solid black lines show the NP simulations, the two dot-dashed lines show PFSS extrapolations relaxed to equilibrium with the imposed outflow, and the dashed line shows the PFSS prediction in the absence of any outflow. Note that all PFSS extrapolations in this plot are taken from the simulated photospheric field. The thick grey line shows the observed daily OMNI2 $|B_x|$ data with a 27-day running average.}
\label{fig:outflow}
\end{figure}

Because it affects only those closed fields already near to the outer boundary, the speed $v_0$ has a relatively small effect on the open flux in the NP model. Physically, this corresponds to the plasma-$\beta$ at lower heights being too low for the plasma flow to deform the magnetic field. This small effect is demonstrated in Figure \ref{fig:outflow}, which shows predicted $B^E$ from various coronal models using the same photospheric field for 5 months in 1999 (along with the observed $B_x$ in grey). The two solid black lines show runs of the NP simulation with different outflow speeds, $v_0=50\,\textrm{km}\,\textrm{s}^{-1}$ (thin line) and $100\,\textrm{km}\,\textrm{s}^{-1}$ (thick line). Halving the outflow velocity is seen only to decrease the level of open flux by about 10\%. The difference is visible in equilibrium levels are reached in the first few days of the simulation, before any bipoles emerge (the initial conditions are a PFSS extrapolation, hence the initial increase). Another effect seen in Figure \ref{fig:outflow} is a right-ward shift of the jagged peaks in open flux, which correspond to flux rope ejections, in the run with lower outflow speed. This is because the outflow affects the evolution of flux ropes after they lose equilibrium (though not before). With lower outflow they take longer to pass through the outer boundary.

To further isolate the effect of the outflow, we have taken PFSS extrapolations at several times, inserted these as initial conditions in the non-potential code, and relaxed to equilibrium with an imposed outflow (but no flux emergence or surface transport). Two such sequences, with $v_0=50\,\textrm{km}\,\textrm{s}^{-1}$ and $v_0=100\,\textrm{km}\,\textrm{s}^{-1}$, are shown by the dot-dashed lines in Figure \ref{fig:outflow}. While there is an enhancement of $30\%$--$40\%$ relative to the original PFSS extrapolation (dashed line), the open flux is still much lower than in the full NP simulation, which better matches the observed IMF. The radial outflow is therefore not responsible for most of this improved agreement. Other effects are important and are considered in the next section.

\section{Open Flux in the Non-potential Model} \label{sec:np}

The total open flux in the NP model originates from three main sources, in addition to a small dependence on the outflow boundary condition as described in Section \ref{sec:outflow}:
\begin{enumerate}
\item The photospheric flux distribution, which originates in the observational magnetogram input.
\item A long-term enhancement to the mean level of open flux, due to inflation by coronal currents.
\item Fluctuations super-imposed on the mean enhancement, caused by the ejection of magnetic flux ropes.
\end{enumerate}
The first source also applies to PFSS extrapolations and was described in Section \ref{sec:poten}. Here we consider the other two sources in turn.

\subsection{Long-term Enhancement by Coronal Currents} \label{sec:currents}

This is a consequence of the basic tendency of magnetic flux tubes to expand as they are stressed by currents \citep{mackay2002b, sturrock1994}. The early numerical force-free model of \citet{barnes1972} showed that field lines expand outwards as the magnetic field is twisted up. In a comparison between global MHD and PFSS solutions, \citet{riley2006} found that the magnetic field in the MHD solution was more inflated and opened, with a realistic ``cusp-like'' morphology of coronal streamers not present in the PFSS solution. Such features are also seen in our NP simulations (Figure \ref{fig:np}).

\begin{figure}
\noindent\includegraphics[width=20pc]{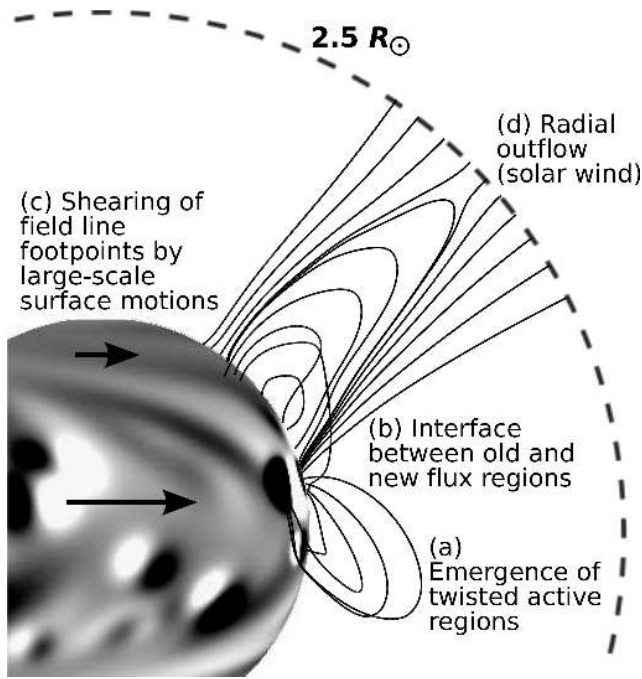}
\caption{The four main sources of coronal currents in our NP model.}
\label{fig:cartoon}
\end{figure}

\begin{figure}
\noindent\includegraphics[width=20pc]{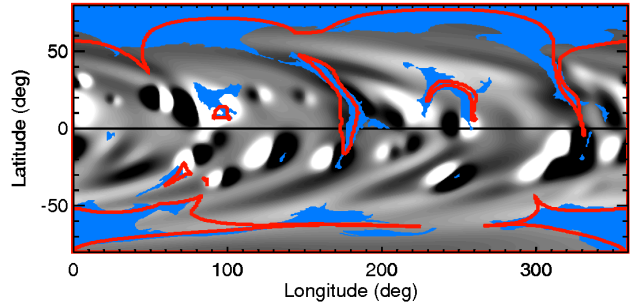}
\caption{Photospheric footpoints of open field lines on 27 October 2000, in the NP model (blue) and PFSS extrapolation (from the simulated surface field, red outlines). Grey shading shows the radial magnetic field in the photosphere (white positive, black negative).}
\label{fig:locations}
\end{figure}

In our NP model there are three main sources of currents, in addition to the radial outflow discussed above. These are shown schematically in Figure \ref{fig:cartoon} \citep[see also][]{yeates2008b}. Firstly, the new bipoles emerge with a non-zero twist (a). This is initially concentrated low down in the center of the bipole, shearing field lines that cross the central polarity inversion line. Secondly, when bipoles emerge they displace pre-existing fields and produce currents at the interfaces between old and new flux systems (b). This can lead to the formation of observed ``intermediate'' filaments at these locations \citep{gaizauskas1997,wang2007,mackay2008}. Thirdly, over days to weeks, surface motions, primarily differential rotation, shear the coronal field, generating further currents (c). Supergranular diffusion converges magnetic flux and helicity toward polarity inversion lines, leading to flux cancellation and the formation of twisted, current-carrying magnetic flux ropes \citep{yeates2009b}. Figure \ref{fig:locations} shows that these different sources of current in the NP model lead to additional open field lines at various locations on the solar surface. The lower latitude open regions that were present in the PFSS model (red outlines) are much extended and their shapes modified, while additional open regions are found in a number of active regions. However, shearing by differential rotation at higher latitudes has also led to extension of the polar coronal holes. We do not discuss these spatial differences in the distribution of open field lines any further in this paper, concentrating instead on the global, integrated open flux over time. In particular, spatial comparison of the NP model with observed coronal hole boundaries is left to future work.

\begin{figure}
\noindent\includegraphics[width=20pc]{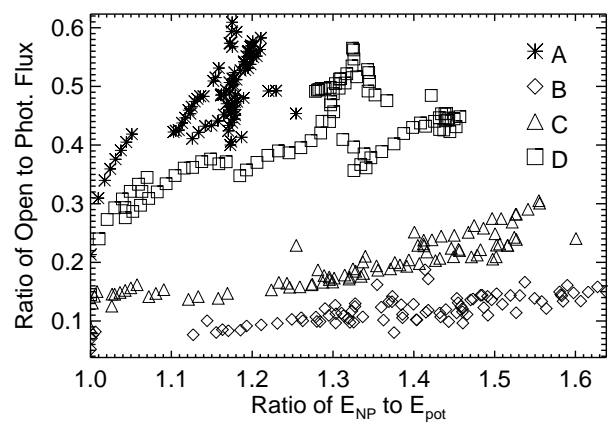}
\caption{Dependence of the ratio of open flux on the ratio of NP to PFSS energy, for simulation periods A to D. Data points are plotted for every second day of each simulation. The ratio of open flux is defined as the ratio of total unsigned radial flux through the source surface $r=2.5R_\odot$ to that through the photospheric boundary $r=R_\odot$.}
\label{fig:currentvsflux}
\end{figure}

Figure \ref{fig:currentvsflux} demonstrates how the fraction of open flux ({\it i.e.}, the ratio of unsigned open flux to total unsigned photospheric flux) depends on the ratio of total magnetic energy in the NP and PFSS models. This energy ratio is a proxy for the total current and measures the ``non-potentiality'' of the global field. Each symbol represents a single snapshot, and the shapes indicate the four simulation periods as in the legend. Note that each run was initialized with a PFSS extrapolation, so initially has an energy ratio of unity. In each simulation run, there is a clear trend for a higher ratio of open flux at times with greater free magnetic energy, consistent with the idea of inflation by the presence of currents. Another clear trend in Figure \ref{fig:currentvsflux} is the solar cycle dependence, which manifests itself in two ways: energy ratios are typically lower at Minimum, but the ratio of open flux to photospheric flux is higher. Since the free energy and current arise from the emergence and interaction of strong active regions, it is no surprise that there is less free energy at Minimum. Indeed, the total parallel current in the simulations increases from period A to period B by about a factor 5.5. The higher fraction of open flux at Minimum is a well known result, arising because at Minimum the open flux decreases much less than the total photospheric flux \citep{wang2000}, owing to the presence of the polar coronal holes. We stress, however, that the absolute amount of open flux ({\it i.e.}, without normalizing by the photospheric flux) is higher at Maximum, in accordance with IMF observations, as is evident in Figure \ref{fig:np}. It is this activity-dependent enhancement of the magnitude of open flux that is the key difference between our NP model and the PFSS model, responsible for the improved match with observations.

\subsection{Fluctuating Enhancement by Flux Rope Ejections} \label{sec:ejections}

\begin{figure*}
\noindent\includegraphics[width=39pc]{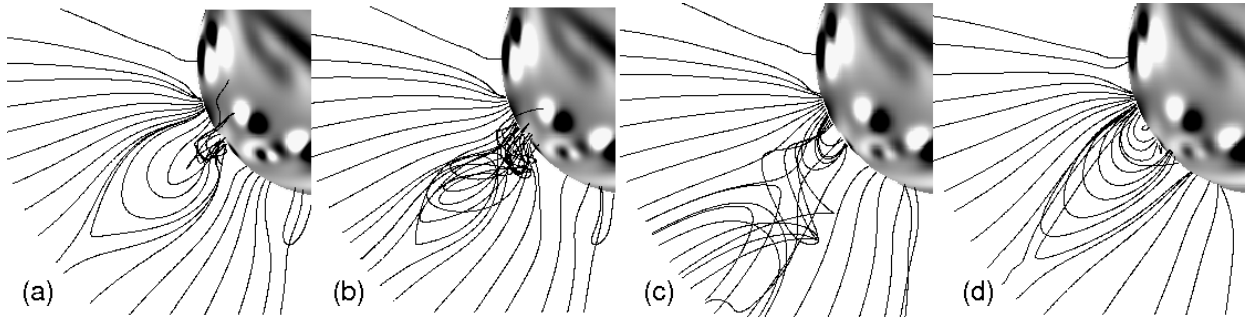}
\caption{Sequence showing the loss of equilibrium of a magnetic flux rope in the NP simulation (period B). The interval between panels (a) and (b) is 4 days, with subsequent panels at 2-day intervals. Grey-shading shows radial magnetic field on the photosphere (white positive, black negative, with a saturation level of   $20\,\textrm{G}$ or $2\,\textrm{mT}$), while selected coronal magnetic field lines are traced from the plane of the sky.}
\label{fig:rope}
\end{figure*}

In addition to the long-term inflation of the magnetic field by coronal currents, there are additional temporary enhancements of the NP open flux lasting for a few days. Many of these are caused by the ejection of flux ropes, which form by flux cancellation at polarity inversion lines and lose equilibrium if they gain too much axial flux relative to the overlying field. An example is shown in Figure \ref{fig:rope}. Using the same magneto-frictional model as in the present paper, \citet{mackay2006b} looked in detail at the flux rope that formed and lost equilibrium between two bipoles in a simplified configuration. They found that the open flux of the two bipoles increased temporarily by a factor $\sim2$ during the ejection as closed field lines were opened and reconnected.

\begin{figure}
\noindent\includegraphics[width=20pc]{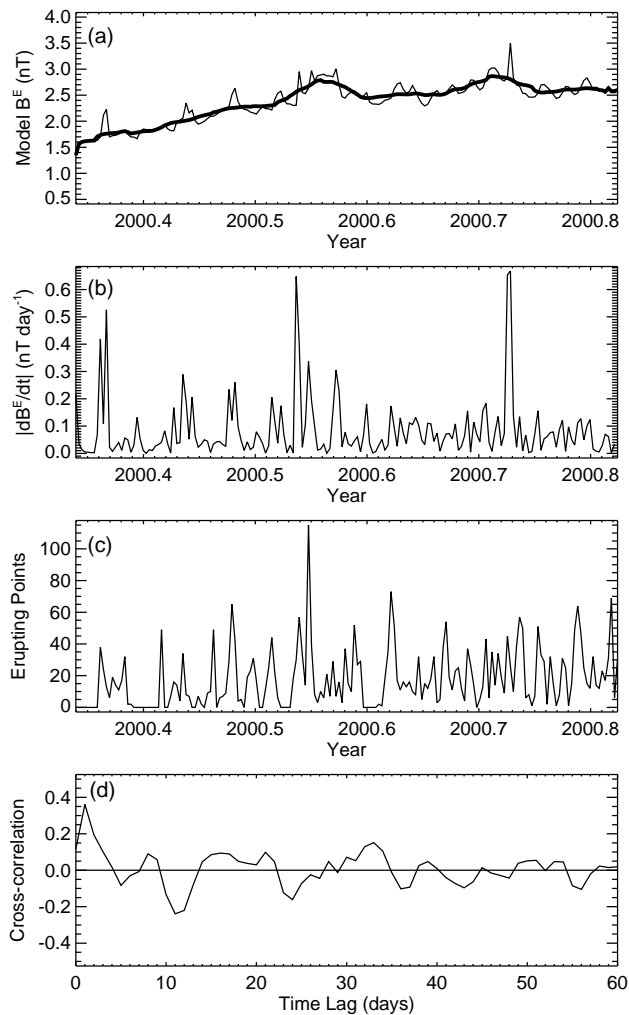}
\caption{Relation between the open flux and flux rope lift-offs in the NP model, for simulation period B. Panel (a) shows the total $B^E$ (thin line), with the ``mean'' value after smoothing with a 14-day running average superimposed (thick line). Panel (b) shows the absolute time rate of change of the (total) $B^E$, while panel (c) shows the number of erupting flux rope points present on each day (see text). Panel (d) shows the cross-correlation between (c) and the fluctuating component in (a).}
\label{fig:liftoffs}
\end{figure}

An analysis of the fluctuations in one of our global simulations (period B) is presented in Figure \ref{fig:liftoffs}. Figure \ref{fig:liftoffs}(a) shows the total $B^E$ (thin line) with a 14-day running mean overlaid in the thick line. This mean component arises from the various longer-term contributions as discussed earlier. Figure \ref{fig:liftoffs}(b) shows the (absolute) time rate of change of $B^E$, where several sharp peaks are clearly visible. To relate these fluctuations to flux rope ejections in the model, Figure \ref{fig:liftoffs}(c) shows the number of ``erupting points'' on each day in the simulation. These are points on a computational grid which (a) are identified as part of a flux rope structure, and (b) have a radial velocity in the magneto-frictional code in excess of $0.5\,\textrm{km}\,\textrm{s}^{-1}$. Details of our automated techniques to detect flux ropes and their ejections in the global simulations are given by \citet{yeates2009b}. We see from Figure \ref{fig:liftoffs}(c) that the number of grid points involved in flux rope ejections has a sharply-peaked structure similar to that of $|dB^E/dt|$. These peaks correspond to the times of flux rope ejections, though some ejections take place gradually over a number of days, and there may be several in progress at one time. There is a reasonable correspondence between peaks in Figure \ref{fig:liftoffs}(b) and \ref{fig:liftoffs}(c)---for example around $2000.54$---although this is by no means one-to-one, demonstrating the complex nature of the global model. Computing the cross-correlation between the number of erupting points and the fluctuating component of $B^E$ (defined as the total minus the mean shown in Figure \ref{fig:liftoffs}a), we find a peak correlation of $0.35$ at a time lag of 1 day, with no strong correlation at later times (Figure \ref{fig:liftoffs}d). This supports our assertion that flux ropes enhance the open flux while they are ejected.

Comparing the different NP runs, we find that ejections occur at the rate of $0.11$, $0.87$, $0.49$ and $0.19$ per day in the final 100 days of runs A, B, C, D respectively. We use only the last 100 days here to avoid the initial ramp-up in the ejection rate (and also the open flux). This solar cycle variation in the number of flux rope ejections leads to a similar variation in contributions to the open flux. This may be seen in Figure \ref{fig:np}, where the predicted $B^E$ for periods B and C shows more short-term fluctuations than that for periods A and D. This is in part responsible for the enhancement of open flux at Maximum which brings the NP model closer to the observed $B^E$ than the PFSS model, although the background inflation by currents is also important. \citet{mackay2006b} estimated that if flux rope ejections occur once every 20 days from each active region, and enhance the open flux for 2 days, then this fluctuating contribution can account for only about $10\%$ of the open flux.

Observationally, flux rope ejections are likely to be manifested as observed coronal mass ejections (CMEs), and indeed CMEs have been linked to freshly opening regions of coronal magnetic flux \citep{luhmann1998}. \citet{owens2008b} recognized the temporary contribution of CMEs to the IMF, and proposed that the IMF is the combination of a constant open flux ``floor'' and transient contributions from Earth-directed CMEs. They found a strong correlation between the IMF strength and Carrington rotation averages of the observed CME rate, and went on to estimate that if each CME contributes $\sim 10^{13}\,\textrm{Wb}$ of magnetic flux to the IMF, it must do so for a period of 30--50 days in order to explain the observed cycle variation of the IMF. While this flux per ejection is comparable in order of magnitude to ejections in our NP model, the timescale for contribution is much longer. By adding the estimated CME contribution to the open flux predicted from the PFSS model, \citet{riley2007} was able to match the observed IMF with a more realistic timescale of about 2 days. These are only rough estimates, however. Our NP simulations represent the first step toward a time-dependent coronal model that self-consistently initiates CMEs and the associated open flux enhancement, although predicting the enhancement by individual observed events is not yet possible. Primarily due to the low time resolution of the magnetogram input to the simulations, the rate of ejections in the model is only about $20\%$ that of observed CMEs, and the model is unable to reproduce the rapid onset of multiple CMEs in individual active regions \citep{yeates2010}. One might initially expect that a more finely-detailed model that produced the observed rate of ejections would yield too high an open flux. However, this may be balanced by two factors. Firstly, many of the additional CMEs recur in the same active region within a few hours to days, so might not generate significant additional open flux. Secondly, the timescale for open flux enhancement by each ejection, which is often a number of days in our simulations, could be too long. More frequent but more rapid ejections may not lead to additional open flux beyond our model. Future work will need to consider the dynamical effects controlling the ejection of flux ropes after they lose equilibrium, which are not taken into account in the magneto-frictional model.

\section{Conclusion} \label{sec:conclusion}

We have studied the open magnetic flux produced by a non-potential (NP) magnetic field evolution model for the lower solar corona (below $2.5R_\odot$). As observational input, this model uses the same observed photospheric magnetograms as the widely used PFSS model, but while the PFSS open flux is determined entirely by the photospheric flux distribution, the NP model predicts additional open magnetic flux due to inflation of the magnetic field by coronal electric currents. These currents arise naturally from the emergence of twisted active regions, their interaction, and their transport over the solar surface by photospheric motions. The total current varies over the solar cycle, leading to greater enhancement of the PFSS open flux at Maximum than at Minimum, as demonstrated in Figure \ref{fig:np}. We propose that the presence of currents in the corona is a viable explanation for the activity-dependent discrepancy between PFSS predictions and {\it in situ} IMF observations at 1AU.

Of course, the influence of coronal currents as modeled here is only one possible explanation for the shortfall in PFSS open flux. To accurately determine the amount of open flux enhancement caused by coronal currents as opposed, for example, to kinematic effects in the solar wind \citep{lockwood2009d}, improved observational input is required to better constrain the model. Firstly, discrepancies between PFSS extrapolations demonstrate already the need for accurate calibration of photospheric magnetographs (as seen in Figure \ref{fig:cycle}). Additionally, for the NP model, a key parameter is the amount of electric current that emerges from the solar interior in bipolar active regions. In this paper, the twist of each newly-emerging bipole has been selected at random from a normal distribution approximating the observed overall distribution of active region helicity \citep[][]{pevtsov1995}. Given this approximation, we have been able to compare only the global open flux with average IMF measurements. In particular, we have not considered the spatial distribution of open flux source regions ({\it e.g.}, coronal holes) on the solar surface, which has potentially important implications for acceleration of the solar wind \citep{cranmer2009}. To assess the more detailed impact of electric currents on the localised distribution of open flux will require reliable measurements of currents in individual active regions. Techniques both for extrapolating non-potential fields \citep{derosa2009} and for estimating the rate of helicity injection \citep{demoulin2009} from photospheric observations are under development. This effort will be greatly aided by the recently-launched NASA {\it Solar Dynamics Observatory}, which promises routine high-resolution, high-cadence, vector magnetic field measurements at the photospheric level.

A promising feature of the NP model is that it reduces the dependence of the open flux on the radius of the outer model boundary. In this paper, the boundary has been fixed at $r=2.5R_\odot$, to allow comparison with PFSS extrapolations. However, this is an arbitrary choice made in the PFSS model to optimize agreement with observed coronal structures, and lacks physical motivation \citep{marsch1984,jiang2010}. In the NP model, the magnetic field strength falls off less strongly with radius. We estimate on average that $|\mathbf{B}_0|\sim r^{-2.3}$ near the outer boundary, rather than $r^{-3}$ as in the PFSS model, so that the open flux is nearer to being independent of $R_{SS}$. In future, we hope to remove the need for an artificial boundary altogether by coupling to a more realistic model of the solar wind.

An important effect included in our NP model but not in existing extrapolation models is the ejection of twisted magnetic flux ropes, which form as a natural consequence of the transport of coronal magnetic field line footpoints by surface motions. Their ejection is a natural limiting mechanism for the build-up of currents in the corona \citep{bieber1995}, acting on a timescale much faster than resistive dissipation. Moreover, it has been suggested that the temporary build-up of closed flux in the heliosphere from the resulting CMEs could be responsible for the discrepancy between PFSS extrapolations and IMF measurements \citep{riley2007,owens2008b}. Indeed, in our NP model, flux rope ejections cause fluctuations, on timescales of a few days, in the total open flux. However, our present quasi-static simulations with observational input only once per 27 day Carrington rotation are unable to accurately estimate the timescale and level of enhancement to the Sun's open flux arising from individual CMEs. This will need to be considered in more detail in our future calibration of the NP model. For example, the additional contribution from ejections might counter-balance the reduction in open flux that would result from extending the outer boundary beyond $2.5R_\odot$. Ultimately, observations of CME contributions to the heliospheric magnetic flux could help to constrain the amount of electric current present in the low corona, and hence the most appropriate balance of relaxation and turbulent diffusion in the NP model. However, it is not clear that changing the timescales of these short-term fluctuations will strongly influence the background, mean level of total open flux in the NP model. This underlying enhancement of open flux arises from coronal currents absent in the PFSS model, and is a long-term phenomenon, reflecting the variation of the Sun's magnetic activity over months and years.


%
%
%
%
%
%

%
%
%
%


\begin{acknowledgments}
We thank the referees for constructive suggestions. ARY and DHM acknowledge financial support from the UK/STFC, and ARY was also supported by NASA contract NNM07AB07C to SAO (where much of this work was undertaken). Simulations used the UKMHD parallel computer in St Andrews (SRIF/STFC), and support from a Royal Society research grant to DHM. The visit of JAC to SAO was supported by NASA grant NNX08AW53. Synoptic magnetogram data from NSO/Kitt Peak were produced cooperatively by NSF/ NOAO, NASA/GSFC, and NOAA/SEL, and made publicly accessible on the World Wide Web. The OMNI IMF data were obtained from the GSFC/SPDF OMNIWeb interface at http://omniweb.gsfc.nasa.gov.
\end{acknowledgments}

%
%
%
%
%
%
%
%
%
%





%
%

\end{article}




%
%
%
%
%
%



\begin{thebibliography}{}



\bibitem[{\textit{Altschuler and Newkirk}(1969)}]{altschuler1969}
Altschuler, M. D., and G. Newkirk, Jr. (1969), Magnetic fields and the structure of the solar corona. I: Methods of calculating coronal fields, {\it Sol. Phys.,} \textit{9}, 131--149.

\bibitem[{\textit{Arge et al.}(2002)}]{arge2002}
Arge, C. N., E. Hildner, V. J. Pizzo, and J. W. Harvey (2002), Two solar cycles of nonincreasing magnetic flux, {\it J. Geophys. Res.}, \textit{107}, A101319.

\bibitem[{\textit{Balogh et al.}(1995)}]{balogh1995}
Balogh, A., E. J. Smith, B. T. Tsurutani, D. J. Southwood, R. J. Forsyth, and T. S. Horbury (1995), The heliospheric magnetic field over the south polar region of the Sun, {\it Science}, \textit{268}, 1007-1010.

\bibitem[{\textit{Barnes and Sturrock}(1972)}]{barnes1972}
Barnes, C. W., and P. A. Sturrock (1972), Force-free magnetic-field structures and their role in solar activity, {\it Astrophys. J.}, \textit{174}, 659--670.

\bibitem[{\textit{Beer}(2000)}]{beer2000}
Beer, J. (2000), Long-term indirect indices of solar variability, {\it Space Sci. Rev.}, \textit{94}, 53--66.

\bibitem[{\textit{Bieber and Rust}(1995)}]{bieber1995}
Bieber, J. W. and Rust, D. M. (1995), The escape of magnetic flux from the Sun, {\it Astrophys. J.}, \textit{453}, 911--918.

\bibitem[{\textit{Canfield et al.}(1999)}]{canfield1999}
Canfield, R. C., H. S. Hudson, and D. E. McKenzie (1999), Sigmoidal morphology and eruptive solar activity, {\it Geophys. Res. Lett.}, \textit{26}, 627--630.

\bibitem[{\textit{Canfield et al.}(2007)}]{canfield2007}
Canfield, R. C., M. D. Kazachenko, L. W. Acton, D. H. Mackay, J. Son, and T. L. Freeman (2007), Yohkoh SXT full-resolution observations of sigmoids: structure, formation, and eruption, {\it Astrophys. J. Lett.}, \textit{671}, L81--L84.

\bibitem[{\textit{Cranmer}(2009)}]{cranmer2009}
Cranmer, S. R. (2009), Coronal holes, {\it Living Rev. Sol. Phys.}, \textit{6}, 3.

\bibitem[{\textit{D\'{e}moulin and Pariat}(2009)}]{demoulin2009}
D\'{e}moulin, P. and E. Pariat (2009), Modelling and observations of photospheric magnetic helicity, {\it Adv. Space Res.}, \textit{43}, 1013--1031.

\bibitem[{\textit{DeRosa et al.}(2009)}]{derosa2009}
DeRosa, M. L. et al. (2009), A critical assessment of nonlinear force-free field modeling of the solar corona for active region 10953, {\it Astrophys. J.}, \textit{696}, 1780--1791.

\bibitem[{\textit{Fisk and Zurbuchen}(2006)}]{fisk2006}
Fisk, L. A., and T. H. Zurbuchen (2006), Distribution and properties of open magnetic flux outside of coronal holes, {\it J. Geophys. Res.}, \textit{111}, A09115.

\bibitem[{\textit{Forbes}(2000)}]{forbes2000}
Forbes, T. G. (2000), A review on the genesis of coronal mass ejections, {\it J. Geophys. Res.}, \textit{105}, A10, 23153--23166.

\bibitem[{\textit{Gaizauskas et al.}(1997)}]{gaizauskas1997}
Gaizauskas, V., J. B. Zirker, C. Sweetland, and A. Kovacs (1997), Formation of a solar filament channel, {\it Astrophys. J.}, \textit{479}, 448--457.

\bibitem[{\textit{Gibson and Fan}(2006)}]{gibson2006}
Gibson, S. E., and Y. Fan (2006), Coronal prominence structure and dynamics: A magnetic flux rope interpretation, {\it J. Geophys. Res.}, \textit{111}, A12103.

\bibitem[{\textit{Jiang et al.}(2010)}]{jiang2010}
Jiang, J., R. Cameron, D. Schmitt, and M. Sch\"{u}ssler (2010), Modeling the Sun's open magnetic flux and the heliospheric current sheet, {\it Astrophys. J.}, \textit{709}, 301--307.

\bibitem[{\textit{Lockwood et al.}(1999)}]{lockwood1999}
Lockwood, M., R. Stamper, and M. N. Wild (1999), A doubling of the Sun's coronal magnetic field during the past 100 years, {\it Nature}, \textit{399}, 437--439.

\bibitem[{\textit{Lockwood et al.}(2009a)}]{lockwood2009d}
Lockwood, M., M. Owens, and A. P. Rouillard (2009a), Excess open solar magnetic flux from satellite data: 2. A survey of kinematic effects, {\it J. Geophys. Res.}, \textit{114}, A11104.

\bibitem[{\textit{Lockwood et al.}(2009b)}]{lockwood2009e}
Lockwood, M., A. P. Rouillard, and I. D. Finch (2009b), The rise and fall of open solar flux during the current grand solar maximum, {\it Astrophys. J.}, \textit{700}, 937--944.

\bibitem[{\textit{Luhmann et al.}(1998)}]{luhmann1998}
Luhmann, J. G., J. T. Gosling, J. T. Hoeksema, and X. Zhao (1998), The relationship between large-scale solar magnetic field evolution and coronal mass ejections, {\it J. Geophys. Res.}, \textit{103}, A46585.

\bibitem[{\textit{Mackay and Lockwood}(2002)}]{mackay2002b}
Mackay, D. H., and M. Lockwood (2002), The evolution of the Sun's open magnetic flux - II. Full solar cycle simulations, {\it Sol. Phys.}, \textit{209}, 287--309.

\bibitem[{\textit{Mackay and van Ballegooijen}(2005)}]{mackay2005}
Mackay, D. H., and A. A. van Ballegooijen (2005), New results in modeling the hemispheric pattern of solar filaments, {\it Astrophys. J. Lett.}, \textit{621}, L77--L80.

\bibitem[{\textit{Mackay and van Ballegooijen}(2006a)}]{mackay2006}
Mackay, D. H., and A. A. van Ballegooijen (2006a), Models of the large-scale corona. I. Formation, evolution, and liftoff of magnetic flux ropes, {\it Astrophys. J.}, \textit{641}, 577--589.

\bibitem[{\textit{Mackay and van Ballegooijen}(2006b)}]{mackay2006b}
Mackay, D. H., and A. A. van Ballegooijen (2006b), Models of the large-scale corona. II. Magnetic connectivity and open flux variation, {\it Astrophys. J.}, \textit{642}, 1193--1204.

\bibitem[{\textit{Mackay et al.}(2000)}]{mackay2000}
Mackay, D. H., V. Gaizauskas, and A. A. van Ballegooijen (2000), Comparison of theory and observations of the chirality of filaments within a dispersing activity complex, {\it Astrophys. J.}, \textit{544}, 1122--1134.

\bibitem[{\textit{Mackay et al.}(2002)}]{mackay2002}
Mackay, D. H., E. R. Priest, and M. Lockwood (2002), The evolution of the Sun's open magnetic flux - I. A single bipole, {\it Sol. Phys.}, \textit{207}, 291--308.

\bibitem[{\textit{Mackay et al.}(2008)}]{mackay2008}
Mackay, D. H., V. Gaizauskas, and A. R. Yeates (2008), Where do solar filaments form? Consequences for theoretical models, {\it Solar Phys.}, \textit{248}, 51--65.

\bibitem[{\textit{Marsch and Richter}(1984)}]{marsch1984}
Marsch, E., and Richter, A. K. (1984), Distribution of solar wind angular momentum between particles and magnetic field - inferences about the Alfven critical point from HELIOS observations, {\it J. Geophys. Res.}, \textit{89}, 5386--5394.

\bibitem[{\textit{Martin et al.}(1994)}]{martin1994}
Martin, S. F., R. Bilimoria, and P. W. Tracadas (1994), Magnetic field configurations basic to filament channels and filaments, in {\it Solar Surface Magnetism}, edited by R. J. Rutten and C. J. Schrijver, NATO Advanced Science Institute, pp. 303--338, Kluwer, Dordrecht.

\bibitem[{\textit{Owens et al.}(2008)}]{owens2008b}
Owens. M. J., N. U. Crooker, N. A. Schwadron, T. S. Horbury, S. Yashiro, H. Xie, O. C. St. Cyr, and N. Gopalswamy (2008), Conservation of open solar magnetic flux and the floor in the heliospheric magnetic field, {\it Geophys. Res. Lett.}, \textit{35}, L20108.

\bibitem[{\textit{Pevtsov et al.}(1995)}]{pevtsov1995}
Pevtsov, A. A., R. C. Canfield, and T. R. Metcalf (1995), Latitudinal variation of helicity of photospheric magnetic fields, {\it Astrophys. J.}, \textit{440}, L109--L112.

\bibitem[{\textit{Pulkkinen}(2007)}]{pulkkinen2007}
Pulkkinen, T. (2007), Space weather: terrestrial perspective, {\it Living Rev. Sol. Phys.}, \textit{4}, 1.

\bibitem[{\textit{Riley}(2007)}]{riley2007}
Riley, P. (2007), An alternative interpretation of the relationship between the inferred open solar flux and the interplanetary magnetic field, {\it Astrophys. J.}, \textit{667}, L97--L100.

\bibitem[{\textit{Riley}(2006)}]{riley2006}
Riley, P., J. A. Linker, Z. Miki\'{c}, R. Lionello, S. A. Ledvina, and J. G. Luhmann (2006), A comparison between global solar magnetohydrodynamic and potential field source surface model results, {\it Astrophys. J.}, \textit{653}, 1510--1516.

\bibitem[{\textit{Rust et al.}(2008)}]{rust2008}
Rust, D. M., D. K. Haggerty, M. K.  Georgoulis, N. R. Sheeley, Y.-M. Wang, M. L. DeRosa, and C. J. Schrijver (2008), On the solar origins of open magnetic fields in the heliosphere, {\it Astrophys. J.}, \textit{687}, 636--645.

\bibitem[{\textit{Schatten et al.}(1969)}]{schatten1969}
Schatten, K. H., J. M. Wilcox, and N. F. Ness (1969), A model of interplanetary and coronal magnetic fields, {\it Sol. Phys.}, \textit{6}, 442--455.

\bibitem[{\textit{Schrijver and DeRosa}(2003)}]{schrijver2003}
Schrijver, C. J., and M. L. DeRosa (2003), Photospheric and heliospheric magnetic fields, {\it Sol. Phys.}, \textit{212}, 165--200.

\bibitem[{\textit{Sch\"{u}ssler and Baumann}(2006)}]{schuessler2006}
Sch\"{u}ssler, M. and I. Baumann (2006), Modeling the Sun's open magnetic flux, {\it Astron. Astrophys.}, \textit{459}, 945--953.

\bibitem[{\textit{Schulz et al.}(1978)}]{schulz1978}
Schulz, M., E. N. Frazier, and D. J. Boucher, Jr. (1978), Coronal magnetic-field model with non-spherical source surface, {\it Sol. Phys.}, \textit{60}, 83--104.

\bibitem[{\textit{Sheeley}(2005)}]{sheeley2005}
Sheeley, N. R., Jr. (2005), Surface evolution of the Sun's magnetic field: A historical review of the flux-transport mechanism, {\it Living Rev. Sol. Phys.}, \textit{2}, 5.

\bibitem[{\textit{Smith et al.}(2001)}]{smith2001}
Smith, E. J., A. Balogh, R. J. Forsyth, and D. J. McComas (2001), Ulysses in the south polar cap at solar maximum: Heliospheric magnetic field, {\it Geophys. Res. Lett.}, \textit{28}, 4159--4162.

\bibitem[{\textit{Snodgrass}(1983)}]{snodgrass1983}
Snodgrass, H. B. (1983), Magnetic rotation of the solar photosphere, {\it Astrophys. J.}, \textit{270}, 288--299.

\bibitem[{\textit{Solanki et al.}(2000)}]{solanki2000}
Solanki S. K., M. Sch\"{u}ssler, and M. Fligge (2000), Evolution of the Sun's large-scale magnetic field since the Maunder minimum, {\it Nature}, \textit{408}, 445--447.

\bibitem[{\textit{Sturrock et al.}(1994)}]{sturrock1994}
Sturrock, P. A., S. K. Antiochos, J. A. Klimchuk, and G. Roumeliotis (1994), Asymptotic forms for the energy of force-free magnetic field configurations of translational symmetry, {\it Astrophys. J.}, \textit{431}, 870--872.

\bibitem[{\textit{Su et al.}(2009)}]{su2009}
Su, Y., A. A. van Ballegooijen, B. W.  Lites, E. E. DeLuca, L. Golub, P. C. Grigis, G. Huang, and H. Ji (2009), Observations and nonlinear force-free field modeling of active region 10953, {\it Astrophys. J.}, \textit{691}, 105--114.

\bibitem[{\textit{Svensmark and Friis-Christensen}(1997)}]{svensmark1997}
Svensmark, H. and E. Friis-Christensen (1997), 
Variation of cosmic ray flux and global cloud coverage-a missing link in solar-climate relationships, {\it J. Atmos. Sol. Terr. Phys.}, \textit{59}, 1225--1232.

\bibitem[{\textit{Ulrich}(1992)}]{ulrich1992}
Ulrich, R. K. (1992), Analysis of magnetic fluxtubes on the solar surface from observations at Mt. Wilson of A5250; A5233, in {\it ASP Conf. Ser. 26, Cool Stars, Stellar Systems, and the
Sun}, edited by M. S. Giampapa and J. A. Bookbinder, pp. 265--267, ASP, San Francisco.

\bibitem[{\textit{van Ballegooijen and Martens}(1989)}]{vanballegooijen1989}
van Ballegooijen, and P. C. H. Martens (1989), Formation and eruption of solar prominences, {\it Astrophys. J.}, \textit{343}, 971--984.

\bibitem[{\textit{van Ballegooijen et al.}(2000)}]{vanballegooijen2000}
van Ballegooijen, A. A., E. R. Priest, and D. H. Mackay (2000), Mean field model for the formation of filament channels on the Sun, {\it Astrophys. J.}, \textit{539}, 983--994.

\bibitem[{\textit{Vieira and Solanki}(2010)}]{vieira2010}
Vieira, L. E. A. and S. K. Solanki (2010), Evolution of the solar magnetic flux on time scales of years to millenia, {\it Astron. Astrophys.}, \textit{509}, A100.

\bibitem[{\textit{Wang and Muglach}(2007)}]{wang2007}
Wang, Y.-M., and K. Muglach (2007), On the formation of filament channels, {\it Astrophys. J.}, \textit{666}, 1284--1295.

\bibitem[{\textit{Wang and Sheeley}(1995)}]{wang1995}
Wang, Y.-M., and N. R. Sheeley, Jr. (1995), Solar implications of ULYSSES interplanetary field measurements, {\it Astrophys. J.}, \textit{447}, L143--L146.

\bibitem[{\textit{Wang and Sheeley}(2002)}]{wang2002}
Wang, Y.-M., and N. R. Sheeley, Jr. (2002), Sunspot activity and the long-term variation of the Sun's open magnetic flux, {\it J. Geophys. Res.}, \textit{107}, A101302.

\bibitem[{\textit{Wang et al.}(1996)}]{wang1996}
Wang, Y.-M., S. H. Hawley, and N. R. Sheeley, Jr. (1996), The magnetic nature of coronal holes, {\it Science}, \textit{271}, 464--469.

\bibitem[{\textit{Wang et al.}(2000)}]{wang2000}
Wang, Y.-M., J. Lean, and N. R. Sheeley, Jr. (2000), The long-term variation of the Sun's open magnetic flux, {\it Geophys. Res. Lett.}, \textit{27}, 505--508.

\bibitem[{\textit{Yang et al.}(1986)}]{yang1986}
Yang, W. H., P. A. Sturrock, and S. K. Antiochos (1986), Force-free magnetic fields - the magneto-frictional method, {\it Astrophys. J.}, \textit{309}, 383--391.

\bibitem[{\textit{Yeates and Mackay}(2009)}]{yeates2009b}
Yeates, A. R., and D. H. Mackay, (2009), Initiation of coronal mass ejections in a global evolution model, {\it Astrophys. J.}, \textit{699}, 1024--1037.

\bibitem[{\textit{Yeates et al.}(2007)}]{yeates2007}
Yeates, A. R., D. H. Mackay, and A. A. van Ballegooijen (2007), Modelling the global solar corona: Filament chirality observations and surface simulations, {\it Sol. Phys.}, \textit{245}, 87--107.

\bibitem[{\textit{Yeates et al.}(2008a)}]{yeates2008a}
Yeates, A. R., D. H. Mackay, and A. A. van Ballegooijen (2008a), Modelling the global solar corona II: Coronal evolution and filament chirality comparison, {\it Sol. Phys.}, \textit{247}, 103--121.

\bibitem[{\textit{Yeates et al.}(2008b)}]{yeates2008b}
Yeates, A. R., D. H. Mackay, and A. A. van Ballegooijen (2008b), Evolution and distribution of current helicity in full-Sun simulations, {\it Astrophys. J.}, \textit{680}, L165--L168.

\bibitem[{\textit{Yeates et al.}(2010)}]{yeates2010}
Yeates, A. R., G. D. R. Attrill, D. Nandy, D. H. Mackay, P. C. H. Martens, and A. A. van Ballegooijen (2010), Comparison of a global magnetic evolution model with observations of coronal mass ejections, {\it Astrophys. J.}, \textit{709}, 1238--1248.

\bibitem[{\textit{Zhao and Hoeksema}(1995)}]{zhao1995}
Zhao, X., and J. T. Hoeksema (1995), Prediction of the interplanetary magnetic field strength, {\it J. Geophys. Res. A}, \textit{100}, 19--33.

\end{thebibliography}
\end{document}